\documentclass[%
 aip,
 amsmath,amssymb,
 reprint,%
]{revtex4-1}
\usepackage{graphicx}% Include figure files
\usepackage{dcolumn}% Align table columns on decimal point
\usepackage{bm}% bold math
\usepackage[utf8]{inputenc}
\usepackage[T1]{fontenc}
\usepackage{mathptmx}
\usepackage{etoolbox}
\usepackage{textcomp}
\usepackage{xcolor}

\newcommand*{\citen}[1]{%
  \begingroup
    \romannumeral-`\x % remove space at the beginning of \setcitestyle
    \setcitestyle{numbers}%
    \cite{#1}%
  \endgroup   
}

\mathchardef\mhyphen="2D
\makeatletter
\def\@email#1#2{%
 \endgroup
 \patchcmd{\titleblock@produce}
  {\frontmatter@RRAPformat}
  {\frontmatter@RRAPformat{\produce@RRAP{*#1\href{mailto:#2}{#2}}}\frontmatter@RRAPformat}
  {}{}
}%
\makeatother

\begin{document}

\preprint{AIP/123-QED}

\title{Thermally Stable Peltier Controlled Vacuum Chamber for Electrical Transport Measurements}

\author{S. F. Poole}
\affiliation{ 
School of Physics and Astronomy, University of Nottingham, University Park, Nottingham, NG7 2RD, United Kingdom
}

\author{O. J. Amin}
\affiliation{ 
School of Physics and Astronomy, University of Nottingham, University Park, Nottingham, NG7 2RD, United Kingdom
}

\author{A. Solomon}
\affiliation{ 
School of Physics and Astronomy, University of Nottingham, University Park, Nottingham, NG7 2RD, United Kingdom
}
\author{L. X. Barton}
\affiliation{ 
School of Physics and Astronomy, University of Nottingham, University Park, Nottingham, NG7 2RD, United Kingdom
}

\author{R. P. Campion}
\affiliation{ 
School of Physics and Astronomy, University of Nottingham, University Park, Nottingham, NG7 2RD, United Kingdom
}

\author{K. W. Edmonds}
\affiliation{ 
School of Physics and Astronomy, University of Nottingham, University Park, Nottingham, NG7 2RD, United Kingdom
}

\author{P. Wadley}
\email{Peter.Wadley@nottingham.ac.uk}
\affiliation{ 
School of Physics and Astronomy, University of Nottingham, University Park, Nottingham, NG7 2RD, United Kingdom
}

\date{\today}

\begin{abstract}
    The design, manufacture and characterisation of an inexpensive,  temperature controlled vacuum chamber with millikelvin stability for electrical transport measurements at and near room temperature is reported. A commercially available Peltier device and high-precision temperature controller are used to actively heat and cool the sample space. The system was designed to minimise thermal fluctuations in spintronic and semiconductor transport measurements but the general principle is relevant to a wide range of electrical measurement applications. The main issues overcome are the mounting of a sample with a path of high thermal conductivity through to the Peltier device and the heat-sinking of said Peltier device inside of a vacuum. A copper slug is used as the mount for a sample and a large copper block is used as a thermal feedthrough before a passive heatsink is used to cool this block. The Peltier device provides 20~W of heating and cooling power achieving a maximum range of 30~K below and 40~K above the ambient temperature. The temperature stability is within 5~mK at all set points with even better performance above ambient temperature. A vacuum pressure of $10^{-8}$~hPa is achievable. As a demonstration, we present experimental results from current-induced electrical switching of a CuMnAs thin film. Transport measurements with and without the Peltier control emphasise the importance of a constant temperature in these applications. The thermal lag between the sample space measurement and the sample itself is observed through magnetoresistance values measured during a temperature sweep.
\end{abstract}

\maketitle

\section{Introduction}
Electrical transport measurements are ubiquitous in materials research, such as in the study of electronic, thermoelectric and spintronic devices \cite{Zelezny_review,MagRoadmap2020,Shao_IEEE2021,sun2021future,sangwan2018electronic,jia2021thermoelectric}. In most cases, these devices are used at or near to ambient conditions, with the device encased in a protective case, such as under potting compound. The characterisation measurements performed in the pursuit of such devices, however, are typically performed in atmosphere, lacking effective temperature control, or in cryostats which often do not perform well near to room temperature. 

Here, the design, manufacture and characterisation of an inexpensive, stable temperature controlled vacuum chamber for electrical transport measurements at and near room temperature are reported. This system uses a Peltier device (also known as a thermo-electric heat pump, thermo-electric generator, or thermo-electric cooler) to control the sample temperature and has a short thermal path between the sample and a thermal reservoir to aid in the removal of waste heat from the sample.

% The accurate control and measurement of temperature in vacuum systems is of crucial importance throughout condensed matter physics research. Despite its importance, few current systems allow for low cost, easy manufacture, and high stability readings at room temperature. The majority of the focus in design and development of such systems is often limited to extreme temperature environments such as cryogenic and high temperature systems, despite the commercial success of products often hinging on near room temperature operation. In this paper we report the design and manufacture of a relatively simple and low-cost desk-top prototype system which places a Peltier device (also known as thermo-electric heat pump, thermo-electric generator, or thermo-electric cooler) inside a vacuum chamber. This system is intended for use in transport measurements such as antiferromagnetic switching, for which stable control near room temperature is desirable. An example of such an experiment in the system is included with associated performance characteristics of the temperature control and vacuum capabilities.\\
A Peltier device is a bi-directional heat pump capable of high precision temperature control within a $\sim50$~K range relative to room temperature. They are therefore a popular solution for temperature control near ambient temperature in many non-vacuum applications\cite{shilpa2023systematic}. A Peltier device can rapidly switch between heating and cooling to achieve stable and high precision temperature control. Only a few other systems have been developed which utilise Peltier devices to control temperature inside vacuum chambers, for example for x-ray \cite{mudd1987temperature} and neutron scattering \cite{bonetti1997very}, and material processing applications \cite{christiansen2020novel}. In Ref.~\citen{raihane2012simple}, Raihane \textit{et al.} report on a system in which a Peltier device, mounted outside a vacuum chamber, is used to control the temperature for measurements of the glass transition temperature in thin film polymers. Their design is effective for their specific requirements, however for transport measurement applications, several features could be optimised. 

In Ref.~\citen{raihane2012simple}, the vacuum chamber is sealed with copper gaskets making swapping samples expensive and difficult; the sample cell is separately sealed introducing more complexity; and the number of electrical connections to the sample is fewer than required for more general transport measurements. Our aim was to create a system that is more generally applicable to condensed matter research and to also include the following performance criteria: pressure $<$ 10$^{-6}$~hPa, stability within $\pm 10^{-2}$~K and range of -5--50~$^\circ$~C. The sample should have up to 12 electrical connections and there must be a good thermal path between the Peltier device and the sample. Additionally, the Peltier power wires should be electrically shielded from the measurement wires to avoid interference.

\begin{figure*}
    \centering
    \includegraphics[width=.9\textwidth]{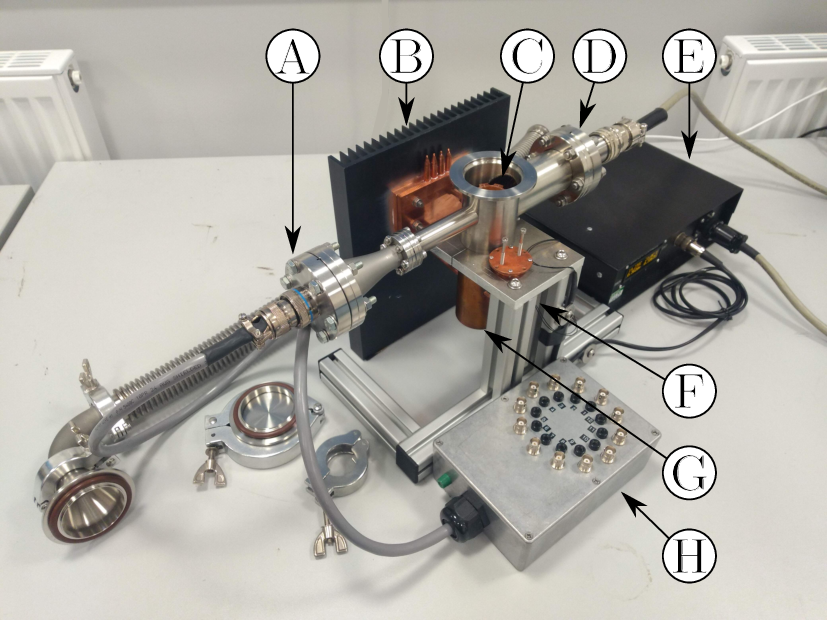}
    \caption{\label{fig:photo} Photograph of the fully assembled system.  The beige cable runs from the power feedthrough (D) to connect the Peltier device and PT100 sensor to the temperature control unit (E) which is in a housing with the mains AC to DC converter, screen and connectors. The black cable connects the external thermocouple (F), which is embedded in the thermal feedthrough (G), to the temperature control unit. The grey cable connects the signal feedthrough (A) to the breakout box (H) which provides convenient access to each of the 12 sample wires using either the central pin of a BNC connector or a 4mm banana plug. The heatsink assembly (B) can be seen and the main chamber (C) which contains the sample space and Peltier device (within the vacuum chamber) is also shown.}
\end{figure*}

The resulting design, photographed in Fig.~\ref{fig:photo}, utilises a large mass of high purity copper as a thermal feedthrough, to which the Peltier device and sample space is attached. This copper is then cooled external to the vacuum using a conventional passive method. The system has since become a staple in our research and so we hope to share our insights with the reader now. In the following, the key design features are explained with some important compromises being highlighted before the performance in terms of vacuum and temperature control are reported. To demonstrate a typical application and highlight the significance of stable temperature control, we present as an example some current-induced switching measurements of an antiferromagnetic CuMnAs thin film. Finally some suggestions and possible improvements are provided for others wishing to produce a similar system who may have slightly different requirements or expectations.

\section{Design}

\begin{figure*}
    \centering
    \includegraphics[width=.9\textwidth]{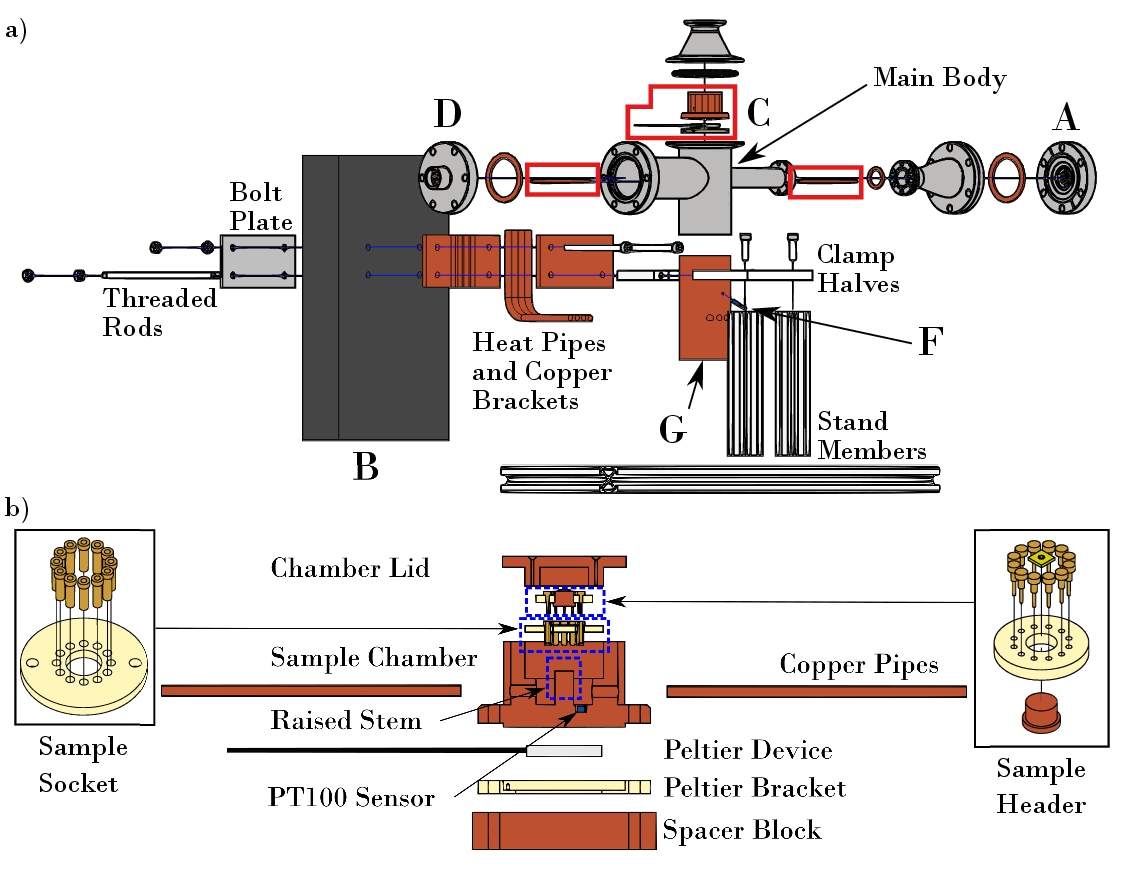}
    \caption{\label{fig:Explodeds} Exploded views of the 3D CAD drawings for the system excluding wiring and most fasteners. a) shows the assembly of the whole system while b) focuses on the smaller parts of the sample space. The letter labels match those in Figure~\ref{fig:photo}: A is the signal feedthrough, B is the heatsink, C is the sample space sub-assembly, D is the power feedthrough, F is the external thermocouple and G is the thermal feedthrough. The red boxes in (a) correspond to the components drawn in panel (b).}
\end{figure*}

The system comprises a T-shaped vacuum chamber, as shown in the computer aided design (CAD) drawing of Fig.~\ref{fig:Explodeds}(a), with a pumping port on top and a sealed bottom achieved by vacuum brazing a copper cylinder inside it. This copper cylinder acts both as a mounting point and as a thermal feedthrough. The two arms of the T are used for signal feedthroughs. The Peltier device is sandwiched between a copper sample space and the bottom copper cylinder of the vacuum chamber, with a bracket to tidy the wiring and keep the Peltier device central. Four holes in the copper cylinder below the vacuum chamber allow L-shaped heat pipes to conduct heat between the Peltier device and heatsinks. The heatsink acts as a heat source for the Peltier device when heating the sample space, and as a radiator to expel waste heat when the device is cooled. A two-part copper bracket sandwiches the pipes. A large passive heatsink was chosen over more common, small, fan assisted heatsinks because the vibrations of a fan may induce additional noise to sensitive measurements. The signal feedthroughs facilitate a 12-pin header on one side and the two power pins of the Peltier device and four pins of a PT100 platinum resistor on the other. The stand was made from extruded aluminium for the base and aluminium plate for the bracket. Stainless steel threaded rod was used to secure the heatsink to the stand and regular nuts and machine screws were used to apply mounting pressure to the copper bracket and heatsink contact areas. A thermocouple is inserted into the copper cylinder as centrally and as close to the Peltier as possible to provide heatsink temperature monitoring.

The sample space houses the resistor and the 12-pin sample header's matching socket as shown in Fig.~\ref{fig:Explodeds}(b). The sample is attached to a copper slug which runs through the center of the sample header, using thermally conductive varnish or silver conductive paste, before bonding wires are used to connect the sample to the pins of the header. The header can then be put into the socket, The lid and the raised stem of the sample space are designed to apply a small amount of pressure onto the header to ensure good thermal contact between the sample and the Peltier device through the raised stem. The spacer block is simply the admission of a miscalculation whereby the pipes could not be inserted unless the sample space was raised by a centimetre. The sample socket is mounted to the sample space using two machine screws made from polyether ether ketone (PEEK), a machinable polymer suitable for high vacuum applications. The sample space, Peltier bracket and space block are all fastened to the copper cylinder of the vacuum chamber using steel screws. 

To control the temperature, an off the shelf 20~W, 20~mm $\times$ 20~mm Peltier device was controlled by a Meerstetter Engineering\textsuperscript{TM} TEC-1089 precision Peltier temperature controller. This reads both the sample space temperature and the external copper block temperature and has a range of features including protections against over-current when changing from heating to cooling and auto-tuning of the proportional integral derivative (PID) parameters. %The provided software is thorough and useful although a little overwhelming. A determined individual can interface with the instrument using any programming language using the provided details of the serial interface, but there are Python and LabView interfaces available from the manufacturer.

\section{Manufacture}

The main body of the vacuum chamber was manufactured by LewVac by welding the tubes and flanges to make the Tee shape and then vacuum brazing the copper cylinder into the large tube. The body is a 50.8~mm outer diameter (OD) stainless steel tube with one 40~mm OD tube and one 19~mm OD tube. The 50~mm tube has a Klein flange for quick and easy access while the other two flanges are copper flanges for a better and more permanent seal. The 19~mm tube is expanded with an adapter to accommodate a 19-pin signal feedthrough and the 40~mm tube has a 6-pin power feedthrough. Four of these pins are used for the PT100 sensor and the remaining two for the Peltier device itself. Any internal thermal contacting face is joined with, vacuum-safe thermally conductive varnish while all external thermal contacts are joined with thermal paste.\\
The in-house workshop's computer numerically controlled (CNC) milling machines and manual lathes are the primary tools for the manufacture of the three custom parts and to drill the four holes in the copper cylinder. These machines are also utilised to manufacture the proprietary headers. All the mounting holes were drilled using the CNC mill and the heatpipe holes in the thermal feedthrough were first drilled and then reamed to enable a press fit.

The sample space and lid were first turned on a lathe before the internal geometries were milled. Through holes were drilled for bolting to the vacuum chamber and holes were drilled and tapped for attaching the lid and sample socket. The Peltier bracket, sample socket body and sample header bodies are made from PEEK. The pins and sockets were press fitted to their respective bodies. The Peltier device is inserted into the bracket with its wires fed through holes in the outer cylindrical face to be fed to the power feedthrough. The copper heatpipe brackets were made by first drilling four holes in a block of copper and then splitting the block in half using a slitting saw on the mill. The meeting face of one bracket and the heatsink were polished, removing the anodising on the heatsink, to maximise thermal conduction. The stand members are made from extruded aluminium and the clamp halves were made by milling a 50.8~mm hole in a piece of 10~mm thick aluminium and then splitting it in twain. One half has a through hole to clear the threaded rods and the other is tapped to a matching thread. 

The wiring was done using vacuum safe solder to connect single core enamelled wires to the 12-pin sample socket and the PT100 sensor. The Peltier power wires were multi-core wires protected with ceramic beads. The feedthroughs have been designed such that the feedthrough can be removed from the wires on the inside and outside. The signal feedthrough uses crimp connectors to make a plug that can be inserted to the feedthrough as one piece. The power feedthrough uses double-ended copper screw terminals to make it possible to disconnect the wires so no solder is needed on these connections. On the outside, mil-spec connectors are typical and so these were used. The Peltier controller and necessary power supply and connectors are all housed in one box. The external thermocouple is press fitted into the copper cylinder with thermal paste to improve thermal conductivity and the thermocouple has its own connector on the controller housing (the thin black cable in Fig.~\ref{fig:photo})

\section{Performance Characterisation}

\begin{figure}
    \centering
    \includegraphics[width=0.45\textwidth]{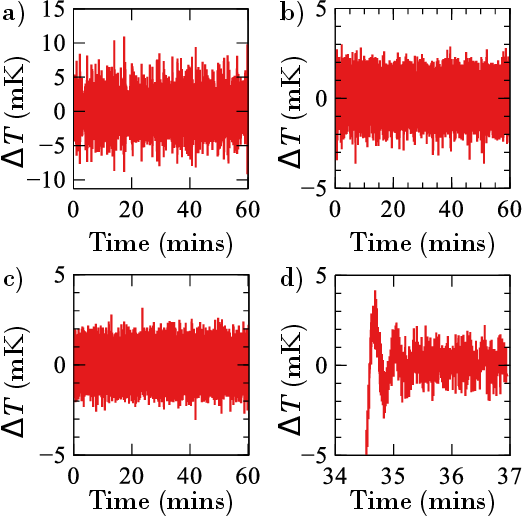}
    \caption{\label{fig:performance} a-c) Plots of thermal stability over time for a) $T=-5$$^\circ$C, b) $T=21$$^\circ$C and c) $T=60$$^\circ$C with a room temperature of around 18~$^\circ$C. d) Stabilisation portion of a ramp from $T=21$$^\circ$C to $T=60$$^\circ$C with a target ramp rate of 0.01~$^\circ$C/s.}
\end{figure}

The key performance criteria are vacuum pressure, temperature control range, thermal stability and sample thermal response. Pressures of $10^{-6}$~hPa are readily achieved just by pumping with a turbo pump. If lower pressures are required, by heating the sample space using nothing but the Peltier device, it is possible to reduce the pressure to $10^{-8}$~hPa. The turbo pump can be run continuously. If vibrations are found to introduce measurement noise then a valve can be closed to isolate the chamber. The leak rate was estimated to be $10^{-6}$~hPa~s$^{-1}$ and so high vacuum will be lost within an hour.

The maximum temperature range was found to be -8~$^\circ$C to +65~$^\circ$C with an ambient temperature of 22~$^\circ$C. The stability was tested and can be seen in Fig.~\ref{fig:performance}~(a-c). The stability was measured at (a) -5~$^\circ$C, (b) 21~$^\circ$C and (c) 60~$^\circ$C, yielding root-mean-square deviations from the setpoint of 1.6~mK, 0.7~mK and 0.7~mK respectively. Figure \ref{fig:performance} (d) shows the amount of overshoot when ramping the temperature from 21~$^\circ$C to 60~$^\circ$C using the ramp rate suggested by the autotuning. The maximum overshoot is 4~mK. The system is fully stable within about a minute after reaching the target temperature. 

\section{application example}

\begin{figure*}
    \centering
    \includegraphics[width=0.84\textwidth]{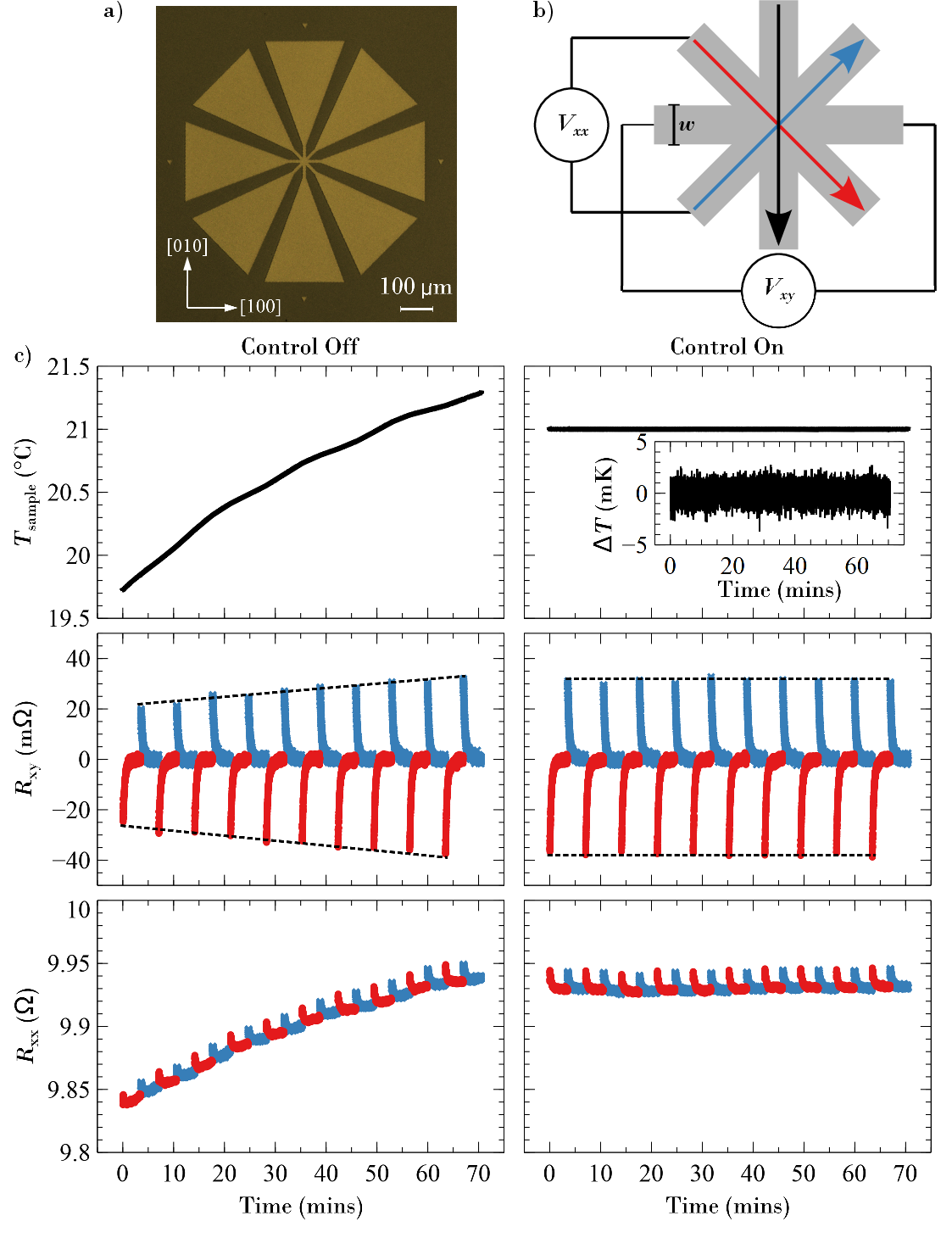}
    \caption{\label{fig:switching} Current-induced resistive switching experiment conducted in the Peltier-controlled system. a) Optical micrograph of the 8-arm CuMnAs device, with an arm width of $w=10$~{\textmu}m. b) Schematic of the pulse and probe geometry with the arm width, $w$, illustrated. 51~mA electrical pulses were applied along the diagonal arms of the device (red and blue arrows corresponding to red circles and blue crosses in c)). After each pulse a continuous probe current was applied along the vertical arm of the device (black arrow) and the longitudinal voltage, $V_{\text{xx}}$, and transverse voltage, $V_{\text{xy}}$, were measured. c) Switching measurements were carried out with the Peltier controller turned off (left column) and turned on with set temperature 21~$^\circ$C (right column). The top two plots show the sample space temperature measured by the PT100 sensor. The middle and bottom plots show the respective $R_{\text{xy}}$ and $R_{\text{xx}}$ measured during a sequence of 10 pairs of pulses. As shown comparing the left and right plots, stable temperature is crucial for achieving reproducible switching data. The size of the $R_{\text{xy}}$ signal increases as the system temperature increases and the base $R_{\text{xx}}$ value follows the temperature drift.}
\end{figure*}

CuMnAs is an antiferromagnetic material in which the magnetic order parameter can be switched using electrical current pulses\cite{wadley2016electrical,Grzybowski2017}. Above a threshold pulse current density, CuMnAs thin films exhibit a resistive switching signal that relaxes to equilibrium over a time-scale of several minutes at room temperature \cite{olejnik2017antiferromagnetic,MatallaWagner2019,Kaspar2021}. The signal amplitude and decay rate display strong temperature dependence, providing an ideal study of the temperature stability in the Peltier controlled transport system.

A room-temperature switching experiment was carried out following the procedures from Refs. \cite{wadley2016electrical, olejnik2017antiferromagnetic, Kaspar2021}. An 8-arm device, with 10~{\textmu}m arm width, was fabricated from a 46~nm layer of CuMnAs epitaxially grown on GaP(001) using established methods\cite{wadley2013tetragonal}. Figure~\ref{fig:switching}(a) shows an optical micrograph of the patterned device. The device was secured onto the sample header using GE varnish and wire bonds were made between the device contact pads and the sample header pins. Once the device was loaded into the system sample socket, situated in the sample chamber as shown in Fig.~\ref{fig:Explodeds}(b), the system was evacuated to $10^{-6}$~hPa pressures.

Current pulses of 51~mA amplitude, 1~ms duration, were applied alternately along the diagonal arms of the device, as shown by the red and blue arrows in Fig~\ref{fig:switching}(b). After each pulse, a continuous 1~mA probe current was applied along the vertical arm of the device and the longitudinal voltage, $V_{\text{xx}}$, and transverse voltage, $V_{\text{xy}}$, were measured for the following $\sim 200$~s. The plots in Fig.~\ref{fig:switching}(c) show the measured sample space temperature (top row), $R_{\text{xy}}$ (middle row), and $R_{\text{xx}}$ (bottom row) during a pulsing sequence conducted with the Peltier controller turned off (left column) and set to 21~$^\circ$C (right column). Note that the $R_{\text{xy}}$ signals are centred on zero by subtracting the mean of the final 100 points in each trace.

Both the $R_{\text{xy}}$ and $R_{\text{xx}}$ signals exhibit appreciable temperature dependence. With the Peltier controller turned off, a sample space temperature drift of $\sim1.5$~$^\circ$C caused a change in the $R_{\text{xy}}$ signal amplitude of $\approx 30$~\%, indicated by the dashed lines, and an $R_{\text{xx}}$ base value drift of 1~\%, following the behaviour of the sample temperature. When the Peltier controller was turned on, a stable sample temperature of $21.000\pm0.003$~$^\circ$C was achieved. The $R_{\text{xy}}$ amplitude was consistent for the complete pulsing set, as shown by the dashed horizontal lines. The $R_{\text{xx}}$ base value was stable within an 0.02~\% variation.

\begin{figure*}
    \centering
    \includegraphics[width=0.84\textwidth]{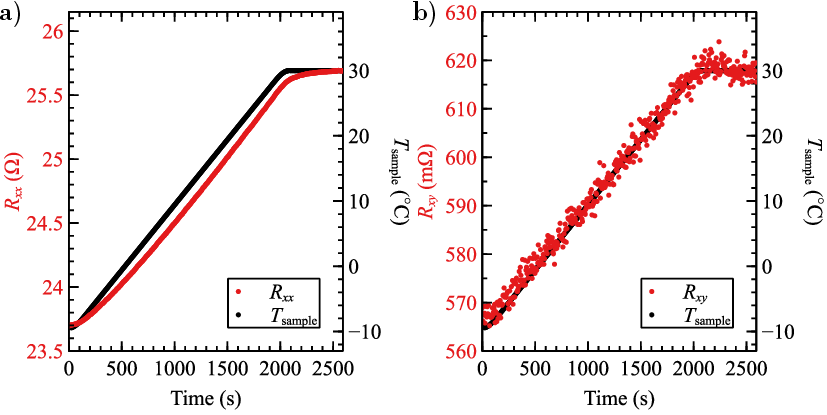}
    \caption{\label{fig:tempsweep} Plots of (a) $R_{xx}$ and (b) $R_{xy}$ for an 8-arm device with $w=5$~{\textmu}m arms fabricated on a 60~nm layer of CuMnAs$_{0.9}$Sb$_{0.1}$ during a temperature sweep from -8--30~$^{\circ}$C over $\approx30$ minutes, with the measurement continuing for a further 500~s after the sweep. $V_{xx}$ and $V_{xy}$ were measured simultaneously such that the resistances were calculated by dividing these values by the applied current of 0.5~mA.}
\end{figure*}

In order to test the thermal conductivity between the sample and the copper chamber, the transverse and linear resistances were measured during a temperature sweep in the system. The 8-arm device used for this study was fabricated from a 60~nm layer of CuMnAs$_{0.9}$Sb$_{0.1}$ epitaxially grown on GaAs(001). The Sb doping of this sample changes the strain but leaves the conductivity and magnetic properties largely unaffected\cite{BartonThesis2023} and so the sample can be considered the same as pure CuMnAs for the sake of this test. A current was applied along the [110] easy axis and the transverse and linear voltages were measured using the geometry shown in Fig.~\ref{fig:switching}(b). Figure~\ref{fig:tempsweep} shows the resulting $R_{xx}$ and $R_{xy}$ values (red points) and temperature readings (black line) as a function of time during the temperature sweep. The difference between the $R_{xx}$ values and the temperature in panel (a) suggests a thermal lag of $\approx 100$~s while the $R_{xy}$ values in panel (b) track the temperature changes almost exactly. The linear resistance value has a direct dependence on the temperature and so this result should be considered to be a more accurate measure of the sample temperature and shows some thermal lag between the sample and the Peltier device readout value. The transverse resistance measurements should not have a large temperature dependence and so this changing value is indicative of mixing between the linear and transverse voltage measurements, as a result of deviations from the perfect device geometry in fabrication, and the delayed response may be hidden by the scatter of the points.

\section{Discussion and Outlook}

A smaller thermal mass, and therefore more rapid thermal response and reduced costs, may appear attractive, however our experiments often include applying high current, short electrical pulses to a sample repeatedly. Because of this, a large thermal mass is preferable to diffuse the generated heat as quickly as possible. It is also helpful simply by increasing the heat capacity and thus reducing the effect of convection on the external side of the Peltier device. It is also for thermal stability reasons that we chose to place the Peltier device inside the vacuum chamber instead of having it exposed to air. Finally, we chose to cool the heatsink passively to avoid vibrations which may introduce noise in sensitive electrical measurements.

It may seem preferable for the system to be designed such that a heatsink could be directly attached to the thermal feedthrough, eliminating the need for heatpipes. This approach, however, necessitates added complexity in either manufacturing the system or changing the sample, since the system would either need to be designed with a complex shape of thermal feedthrough or be upside down to mount the heatsink to (what is currently) the bottom of the thermal feedthrough. This was deemed worse than the chosen solution for this application.

In its current form, a magnetic field can only be applied along one axis (the in plane axis perpendicular to the arms of the tee) but the system cannot be rotated between the poles of a magnet for full in plane measurements because the arms get in the way. The design could be adapted to enable the use of magnetic fields along more than one axis.

With some adaptation, the sample space could be used for many applications requiring this temperature range, however, if there is no need for a large number of signal wires or a small bore, it may be possible to find a simpler solution such as the one used by Raihane \textit{et al.}~\cite{raihane2012simple} There are many configurations possible for a Peltier controlled vacuum system, each with different priorities and drawbacks. In our case, thermal stability, ease of sample swapping and mitigation of electrical noise are our three top priorities in the design of the system.

Overall, this system has provided a relatively cheap and convenient way to ensure minimal thermal fluctuations occur in our transport measurements at and near to room temperature, without the need for cryogens. This was achieved through the combination of a Peltier device placed inside a vacuum chamber, a thermal path almost entirely made of high-purity copper and the used of a high-performance temperature controller. The system allows our samples to be protected from oxidation while under testing and has since become a staple in our research. With some modifications we are confident that most researchers requiring transport measurements would benefit from such a system in their repertoire. The hope is that this design inspires and helps readers looking for a such a solution to discover their ideal system.

\section*{Acknowledgements}
Manufacture and Assembly - Andrew Stuart, Tommy Napier - Engineering Support Team, School of Physics and Astronomy, University of Nottingham, Nottingham UK

\bibliography{RTS}

\end{document}